\documentclass[10pt]{article}
\usepackage[OE]{express}

\begin{document}
\title{Nonlinear optical properties of polycrystalline silicon core fibers from Telecom wavelengths into the mid-infrared spectral region}

\author{H. REN,\authormark{1} L. SHEN,\authormark{1,2,*} D. WU,\authormark{1} O. AKTAS,\authormark{1} T. HAWKINS,\authormark{3} J. BALLATO,\authormark{3} U. J. GIBSON,\authormark{4,5} and A. C. PEACOCK\authormark{1}}

\address{\authormark{1}Optoelectronics Research Centre, University of Southampton, Southampton, SO17 1BJ\\
\authormark{2}Wuhan National Laboratory for Optoelectronics, School of Optical and Electronic Information,
Huazhong University of Science and Technology, Wuhan 430074, Hubei, China\\
\authormark{3}Center for Optical Materials Science and Engineering Technologies (COMSET) and Department of Materials Science and Engineering, Clemson University, Clemson, SC 29634, USA\\
\authormark{4}Department of Physics and Porelabs, Norwegian University of Science and Technology, N-7491 Trondheim, Norway\\
\authormark{5}Department of Applied Physics, KTH Royal Institute of Technology, Stockholm 10044, Sweden}

\email{\authormark{*}L.Shen@soton.ac.uk} %% email address is required


% and  fibers with a few micron-sized core diameters can be fabricated via a post-tapering procedure. %across the two-photon absorption (TPA) edge ($2.25\,\mu\rm{m}$) of silicon
\begin{abstract}
Polycrystalline silicon core fibers (SCFs) fabricated via the molten core drawing (MCD) method are emerging as a flexible optoelectronic platform. Here, the optical transmission properties of MCD SCFs that have been tapered down to a few micrometer-sized core dimensions are characterized from the Telecom band to the mid-infrared spectal regime. The SCFs exhibit low linear losses on the order of a few dB/cm over the entire wavelength range. Characterization of the two-photon absorption coefficient ($\beta_{\rm{TPA}}$) and nonlinear refractive index ($n_2$) of the SCFs reveals values consistent with previous measurements of single crystal silicon materials, indicating the high optical quality of the polysilicon core material. The high nonlinear figure of merit obtained for wavelengths above $2\,\mu$m highlight the potential for these fibers to find application in infrared nonlinear photonics.%  the potential exploitation of these polycrystalline SCFs in nonlinear applications extending beyond telecom band and into the mid-IR regime.
\end{abstract}

  % REPLACE WITH CORRECT OCIS CODES FOR YOUR ARTICLE, MINIMUM OF TWO; Avoid using the OCIS codes for “General” or “General science” whenever possible.
%For a complete list of OCIS codes, visit: https://www.osapublishing.org/oe/submit/ocis/

%%%%%%%%%%%%%%%%%%%%%%% References %%%%%%%%%%%%%%%%%%%%%%%%%

%%%%%%%%%%%%%%%%%%%%%%%%%%  body  %%%%%%%%%%%%%%%%%%%%%%%%%%
\section{Introduction}

Nonlinear silicon photonics has attracted growing interest in the past two decades and  numerous nonlinear effects in silicon materials have been demonstrated for a wide variety of applications ranging from wavelength conversion, signal amplification to broadband supercontinuum generation \cite{Leuthold2010}. Most of these nonlinear effects have been demonstrated in the well-established single crystal silicon-on-insulator (SOI) platform because of its low transmission loss, tight optical confinement and high Kerr nonlinear coefficient $n_{2}$. More recently,  polysilicon (p-Si) waveguides have emerged as an alternative platform, principally as they are much cheaper and more flexible to produce, so can be incorporated into a wider range of geometries \cite{Ballato} and architectures \cite{Orcutt}. However, the high reported propagation losses in waveguides with small, few microns to hundreds of nanometer-sized, dimensions hinder  their use in nonlinear applications.

In complement to the chip-based planar structures, low loss polysilicon waveguides can also be fabricated within the optical fiber platform using the molten core drawing (MCD) method, which is a derivative of the conventional fiber drawing approach \cite{Peacock2016}. However, SCFs fabricated via the MCD method typically exhibit large core sizes (tens of microns) as it is difficult to produce continuous lengths of core with smaller dimensions owing to the high drawing temperatures and speeds \cite{Ballato,Nordstrand}. Thus, a modified tapering procedure has been introduced to scale down the cores of the MCD SCFs to diameters of a few micrometers, or less. Importantly, this post-processing method has also been shown to further improve the crystalline quality of the core material, resulting in a reduction in optical transmission losses in the Telecom band from $12\,\rm{dB/cm}$ in $10\,\mu\rm{m}$ diameter as-drawn fibers, down to $3.5\,\rm{dB/cm}$ in a $1\,\mu\rm{m}$ tapered core fiber \cite{Franz}. This combination of small core size and low loss has allowed for the first observation of nonlinear propagation in a polysilicon material, with the measured nonlinear parameters being comparable with those of single crystal silicon at the Telecom wavelength of $\sim 1.55\mu$m \cite{Fariza}. However, owing to the high nonlinear losses of crystalline silicon in this regime, recently there has been increased interest in their nonlinear properties in the mid-infrared, which is beyond the two-photon absorption (TPA) edge \cite{Liu2010,Zlatanovic2010}. Although measurements of the wavelength dependence of the TPA and Kerr nonlinearity have been reported in bulk materials and SOI waveguides \cite{bristow2007two,lin2007two,Liu2011nonlinear}, as of to date, there have been no reports for polysilicon waveguides and/or for SCFs fabricated via the MCD method.
 %
% previously reported only in bulk single crystal silicon wafers and waveguides in the range of $0.85-2.4\,\mu\rm{m}$ \cite{bristow2007two,lin2007two,Liu2011nonlinear}. For these recently emerged polycrystalline SCFs, a detailed study of their nonlinear parameters for fundamental wavelengths ranging from the telecom to the mid-IR is required and will be a useful guideline for exploiting the nonlinear applications.
%
% in polysilicon  caused by TPA and free-carrier absorption (FCA) in silicon at the telecommunication band limit nonlinear applications which require high pump intensities \cite{yin2007impact}. Therefore, nonlinear silicon photonics research is drifting to the mid-IR region in order to minimize these undesired absorptions \cite{Liu2010,Zlatanovic2010}.

In this paper, we extend the characterizations of the polycrystalline SCFs beyond the telecommunications window and present the first systematic investigation of the transmission properties of this fiber platform spanning from 1.5 to $2.5\,\mu\rm{m}$. This range was specifically selected as it starts from the telecommunications band and continues across the TPA edge $\hbar\omega < E_{{\rm gi}}/2$, where $ E_{{\rm gi}}$ is the indirect bandgap energy of crystalline silicon. A series of wavelength dependent measurements have been conducted using various continuous wave (CW) and short pulse laser sources to determine both the linear losses and the nonlinear transmission properties related to the $\beta_{{\rm TPA}}$ and $n_{2}$  parameters. This characterization provides useful information regarding the quality of our polysilicon core material as well as the dispersion of the FOM$_{{\rm NL}}$ in the vicinity of the TPA edge. By exploiting the low linear losses and the negligible nonlinear absorption for wavelengths beyond $2\,\mu$m, a broad continuum spanning from $1.8\,\mu$m to $3.4\,\mu\rm{m}$ is also generated. The results indicate the potential for SCFs to find use in nonlinear applications across the mid-infrared region where applications include spectroscopy, imagining and sensing.

\section{Fabrication of tapered silicon fibers and experimental setup}
The SCFs studied in this work were fabricated using the MCD technique described in \cite{Nordstrand}.  A thin layer of calcium oxide (CaO) was introduced between the silicon rod and the outer silica cladding to act as a stress buffer, which also prevents oxygen in-diffusing from the cladding to the core during the high temperature draw process. The as-drawn SCFs are polycrystalline, with longitudinal crystal grain sizes on the order of a few millimeters, and have an outer cladding diameter of  $377\,\mu\rm{m}$  and a core diameter of $30\,\mu\rm{m}$.  For nonlinear applications, SCFs with core sizes of a few micrometers or smaller are favored. Thus the as-draw SCFs are subsequently tapered using a Vytran tapering rig (GPX-3400) by slowly feeding the fiber into the hot zone and pulling it from the other end with a higher pulling speed  \cite{Franz}. Various tapered fiber core waists can be produced by controlling the tapering ratio, filament power and pulling velocity. For the following investigations, a uniform waist region with a core diameter of $3\,\mu$m over a $1\,\rm{cm}$ length was selected. 

\begin{figure}[!t]
\includegraphics*[width=\textwidth]{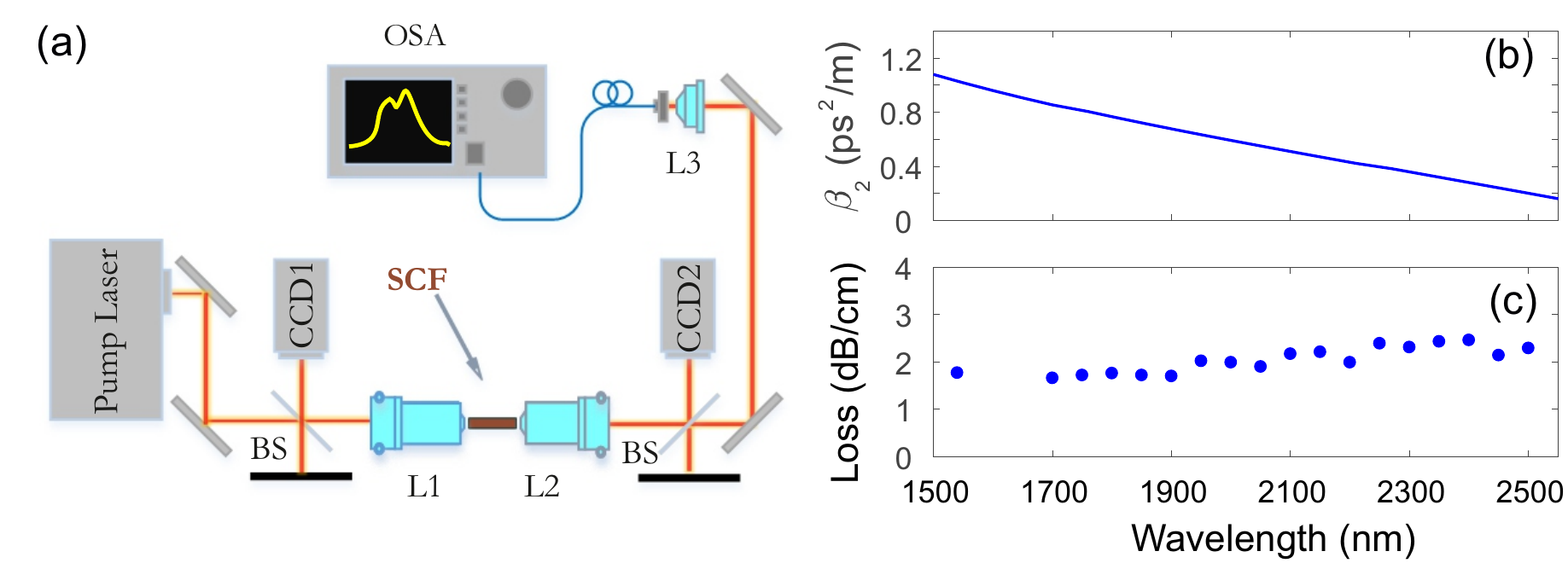}
\caption{(a) Schematic of the transmission setup. Beam-splitter (BS), microscope objective lenses (L1, L2 \& L3), CCD Cameras (CCD1 \& CCD2), optical spectrum analyser (OSA). (b) Calculated dispersion profile of the polycrystalline SCF as a function of wavelength. (c) Linear loss measurements as a function of wavelength.}
\label{omemidfiber_fig1}
\end{figure}

The optical transmission measurements were conducted using the experimental setup shown in Fig. \ref{omemidfiber_fig1}(a).  Two laser sources: (i)  a fiber laser generating hyperbolic secant pulses with a $650\,\rm{fs}$ (FWHM) duration operating at $1.54\,\mu\rm{m}$ with a repetition rate of $40\,\rm{MHz}$  and (ii) a Ti:sappire pumped femtosecond optical parametric oscillator (OPO) for the mid-infrared measurements spanning $1.7-2.5\,\mu\rm{m}$  with a $200\,\rm{fs}$ (FWHM) duration and a repetition rate of $80\,\rm{MHz}$, were employed to measure both the linear losses and nonlinear parameters over the wavelength range $1.5-2.5\,\mu\rm{m}$. Light was coupled into and captured out from the polycrystalline SCF  via free space coupling using two silica microscope objective lens (L1, L2) ($\rm{NA}=0.85$). Then the transmitted light was measured by either a power meter or focused to an optical spectrum analyzer (OSA: Yokagawa AQ 6375) using another microscope objective lens (L3) ($\rm{NA}=0.65$) . The pump power coupled into the polycrystalline SCF was controlled by a variable attenuator (ND filter). CCD cameras were employed to ensure efficient coupling into the center of the core so that the fundamental mode was primarily excited \cite{peacock2012nonlinear}.

As the refractive index for polysilicon is not well documented, the  wavelength dependent group velocity dispersion (GVD) parameter $\beta_2$ of the fundamental mode was estimated by using parameters from the Sellmeier equation for c-Si \cite{Li80}. As shown in Fig. \ref{omemidfiber_fig1}(b), this SCF exhibits normal dispersion across the entire wavelength range: $1.5\,\mu\rm{m}$ to $2.5\,\mu\rm{m}$. This is because the waveguide dispersion of our micron-sized core is not sufficient to compensate for the large normal material dispersion of the high index silicon core. However, this is advantageous for our nonlinear characterization as it ensures that the nonlinear propagation will be governed simply by self-phase modulation (SPM). Furthermore, as the $1\,$cm fiber length is shorter than the dispersion length $L_{{\rm D}}=T_{0}^{2}/|\beta_2|$ at all the wavelengths, we can be sure that the nonlinear effects dominate the high power transmission measurements, i.e., $L_{\rm{NL}}=1/\gamma P_0 \ll L_{{\rm D}}$.

%Although the SCF can support multiple modes in this wavelength region,  a-Si:H core fibers with similar dimensions have been used in our previous reports for nonlinear optical-signal-processing applications and light can be primarily excited into fundamental mode by optimizing the coupling conditions.

\section{Optical transmission properties}

\subsection{Linear propagation loss}
\label{linear}

The linear propagation losses of the SCFs were measured using the cut-back method as described in \cite{lagonigro2010low}. In order to avoid the effects of nonlinear absorption when using the pulsed lasers, the average launch power was kept below $100\,\mu$W. To verify this value, additional measurements were conducted with a $1.55\,\mu\rm{m}$ CW diode, which returned the same loss as our high power fiber laser, confirming that the nonlinear absorption was indeed negligible for these input powers. The loss values stay consistently low at values around $1-2.5\,\rm{dB/cm}$  over the entire wavelength range as shown in Fig. \ref{omemidfiber_fig1} (c).  These loss values are comparable with those of SOI waveguides over the same wavelength region \cite{Rouifed} and present a unique opportunity to investigate the nonlinear properties of this material, and the MCD SCFs. Although the losses measured here are in agreement with our previous work, indicating that the crystallinity of the core is high \cite{Franz}, reducing the losses even further would greatly improve the practicality of these fibers and the material properties are an ongoing study in the tapered SCFs.

%Clearly, the relatively low losses measured for these tapered SCFs suggest that the crystallinity of the core material has been improved and has been verified by Raman spectroscopy in our previous report \cite{Franz}.  

\subsection{Nonlinear absorption in the mid-infrared regime}
\label{Optical limiting of transmission and nonlinear absorption}

\begin{figure}[!t]
\includegraphics*[width=\textwidth]{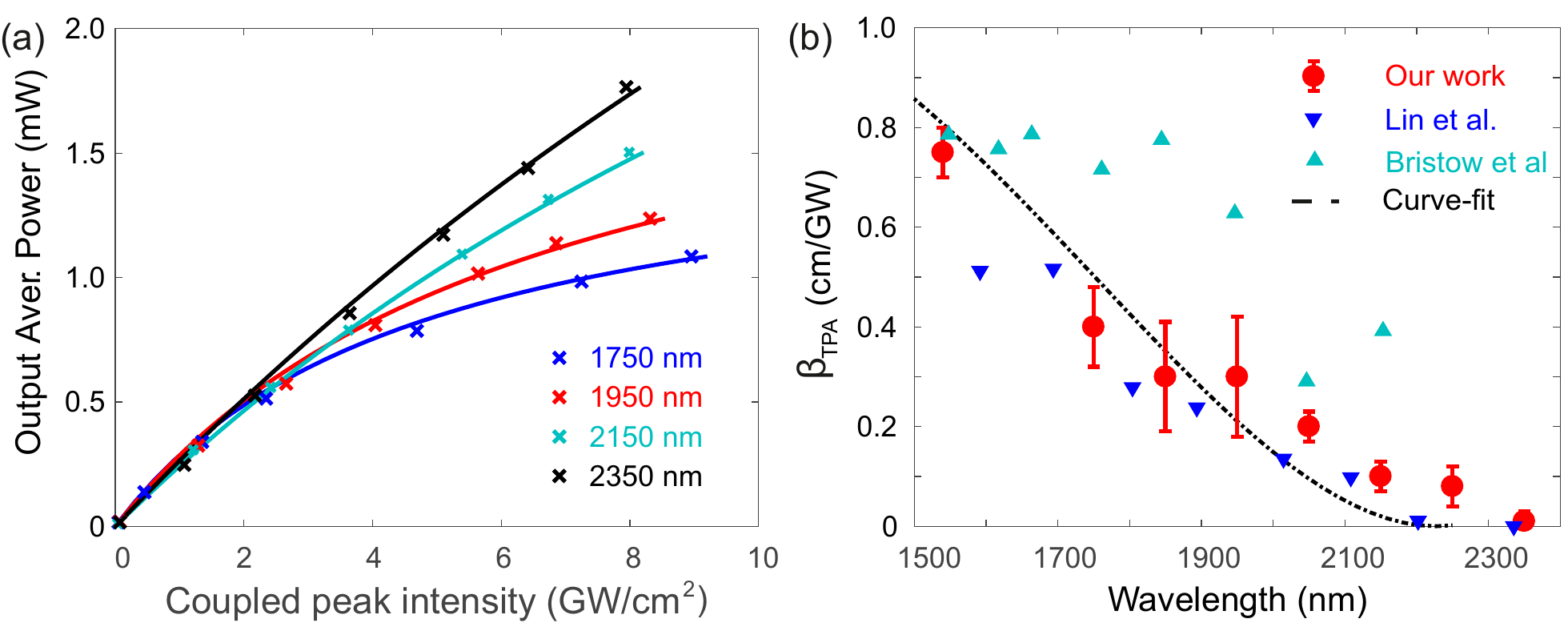}
\caption{(a) Nonlinear absorption measurements for different wavelengths. The solid curves are simulated fits to the data for the corresponding wavelength given in the legend. (b) Measured TPA parameter as a function of wavelength extracted from Fig. \ref{midfiber_fig2}(a), together with data points from previous measurements \cite{lin2007two,bristow2007two} in bulk silicon as labeled. Error bars represent the uncertainty in the input powers.}
\label{midfiber_fig2}
\end{figure}

Characterization of the TPA parameter was performed by measuring the saturation of the output pulses as a function of increasing input power. Our experimental setup employed the same lasers as in Section \ref{linear}, but operating at higher powers for nonlinear measurements. Fig. \ref{midfiber_fig2}(a) plots the results of the output power as a function of coupled input peak intensity for selected pump wavelengths across the TPA window. For all wavelengths up to $2.15\,\mu\rm{m}$, it is obvious that the output powers saturate due to the strong nonlinear absorption caused by TPA when the coupled peak intensity exceeds $ 2\,\rm{GW/cm^2}$. In contrast, the largely linear trend exhibited for transmission at $2.35\,\mu\rm{m}$, which continues up to an input peak intensity of $\sim 8\,\rm{GW/cm^2}$, indicates that TPA is essentially negligible at this wavelength. Although three-photon absorption (3PA) can become significant at wavelengths beyond the TPA edge, the peak intensity in our measurements are insufficient for the observation of the power saturation caused by 3PA \cite{Gholami}.

As discussed in \cite{mehta2010nonlinear}, when dispersion can be neglected, the transmitted pump power through the fiber can be described by a simplified nonlinear Schr\"odinger equation (NLSE) coupled to the rate equation for TPA generated free-carriers:
\begin{eqnarray}
\centering&&\frac{\partial I(z,t)}{\partial z}=-\alpha_lI(z,t)-\beta_{\rm{TPA}}I^2(z,t)-\sigma_{\mathrm{FCA}} N_c(z,t)I(z,t), \label{simplified}
\end{eqnarray}

\begin{eqnarray}
\centering&&\frac{\partial N_c(z,t)}{\partial t}=\frac{\beta_{\rm{TPA}}}{2h\nu_0}{I^2(z,t)}-\frac{N_c(z,t)}{\tau_c}. \label{carrierdensity}
\end{eqnarray}
Here $I(z,t)$, $\alpha_l$, $\sigma_{\mathrm{FCA}}$, $N_c$, and $\tau_c$  represent the pulse intensity, linear loss, free-carrier absorption (FCA) coefficient, free-carrier density, and the carrier lifetime, respectively. To determine the values of $\beta_{\rm{TPA}}$, we fit the experimental data for all wavelengths using the coupled equations, with the remaining material parameters estimated from the single crystal values. The resulting curves are plotted as the solid lines in Fig. \ref{midfiber_fig2}(a), from which we can obtain the TPA parameters presented in Fig. \ref{midfiber_fig2}(b) (red circles). These results show that $\beta_{\rm{TPA}}$ initially drops from $0.75\,\rm{cm/GW}$ down to $0.1\,\rm{cm/GW}$ as the wavelength increases from the Telecom window to the mid-infrared regime ($1.54-2.15\,\mu\rm{m}$), then eventually begins to plateau at a negligible value as the wavelength approaches the edge of the TPA window ($2.25\,\mu\rm{m}$). The $\beta_{\rm{TPA}}$ tends to zero for $\lambda > 2.25\,\mu\rm{m}$, where the sum of the energies of two photons is no longer sufficient to span the bandgap. The trend of decreasing $\beta_{\rm{TPA}}$ seen in Fig. \ref{midfiber_fig2}(b) is what we would expect across this regime, and the results are in good agreement with those reported previously in the literature for bulk silicon \cite{bristow2007two,lin2007two}. The slight differences for the longer wavelengths may stem from the polycrystalline nature of our core material, as there will be material defects at the grain boundaries. However, it is worthwhile to note that our experimental measurements are in very good agreement with the  theoretical fits based on calculations of Garcia and Kalyanaraman \cite{Garcia2006}, as shown in the dashed line in Fig. \ref{midfiber_fig2}(b).

\subsection{SPM induced spectral evolution}

\begin{figure}[!t]
\includegraphics*[width=\textwidth]{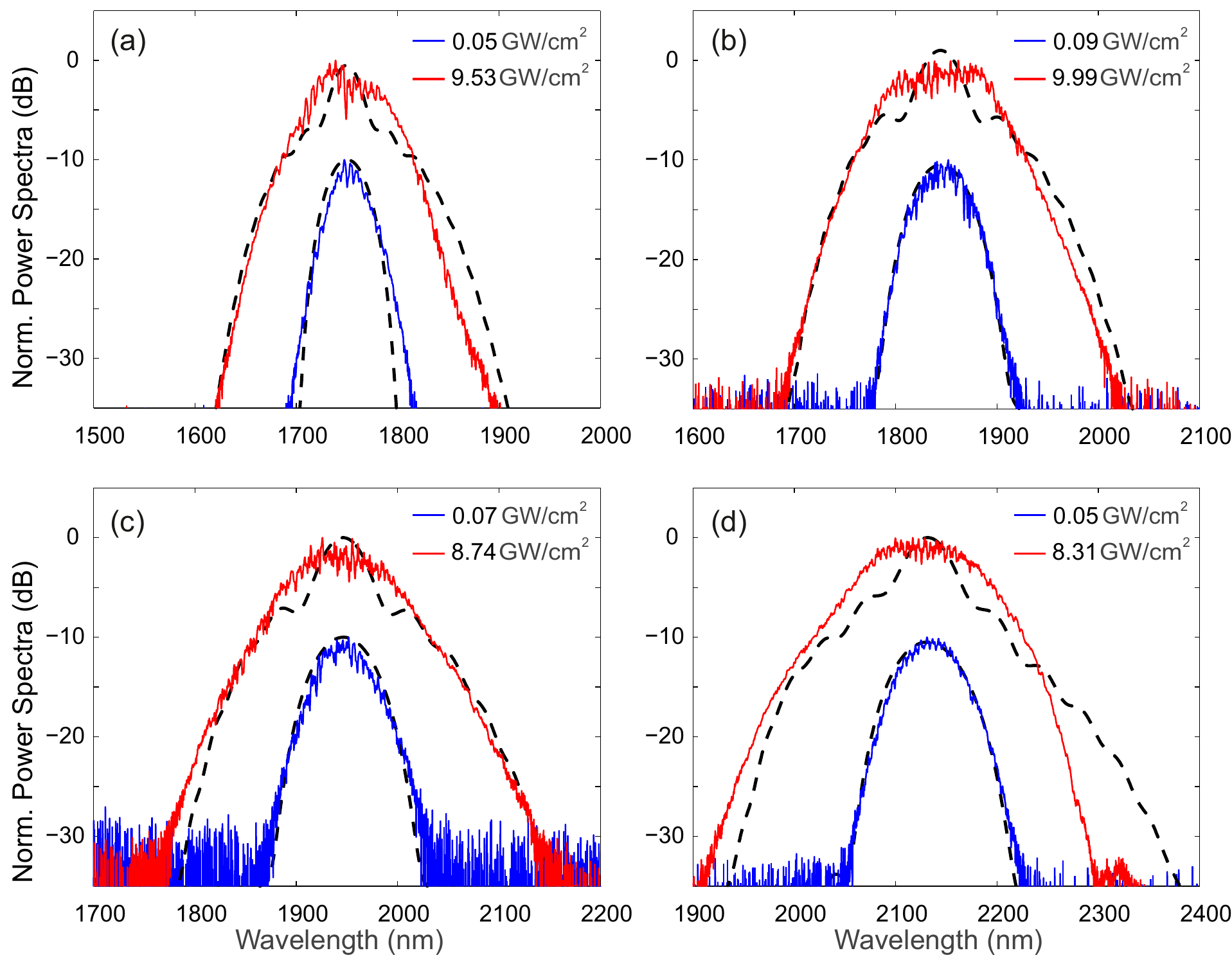}
\caption{Experimental power-dependent transmission spectra as a function of pump center wavelength: (a) $1750\,\rm{nm}$, (b) $1850\,\rm{nm}$, (c) $1950\,\rm{nm}$, and (d) $2150\,\rm{nm}$. The dashed lines are numerical fits obtained by solving the generalized NLSE \cite{mehta2010nonlinear}. The normalized spectra are offset for clarity.}
\label{midfiber_fig3}
\end{figure}

The spectral broadening induced by SPM was then studied to determine the values of the Kerr coefficient $n_2$ over this wavelength range. As in the nonlinear absorption measurements, for these experiments high input peak powers were used, but this time the output spectra of the pulses were monitored on the OSA. The measured SPM spectra are shown in Fig. \ref{midfiber_fig3} for selected wavelengths from $1.55-2.35\,\mu\rm{m}$. For each pump wavelength, the output transmission spectra were recorded at both low and high input peak intensities, as designated by the legends. Here, the results obtained with the low input intensities are essentially free from nonlinear propagation and are included as an indicator of the bandwidth of the input pulse, as a means to determine the size of the initial negative frequency chirp on the pulses, which are not transform-limited. The chirp was not found to vary dramatically over this wavelength range and had a value of $C\sim-1.1$ for all wavelengths \cite{Shen}. The high intensity results are then used to illustrate the strong spectral broadening due to the large Kerr nonlinearity of the silicon core, with bandwidths of more than $300\,\rm{nm}$ obtained for all wavelengths. It is also worth noting that the lack of clear SPM induced modulation on these spectra is due to the initial chirp on the pulses and the noise on the input OPO spectra.

%For comparison, the inset in the top left-hand corner of Fig. \ref{midfiber_fig3} shows the corresponding spectral broadening generated using the $\sim 1\,\rm{ps}$ (FWHM) telecoms fiber laser, where the classic SPM shaping can be clearly observed.

\begin{figure}[!b]
\centering
\includegraphics[width=\textwidth]{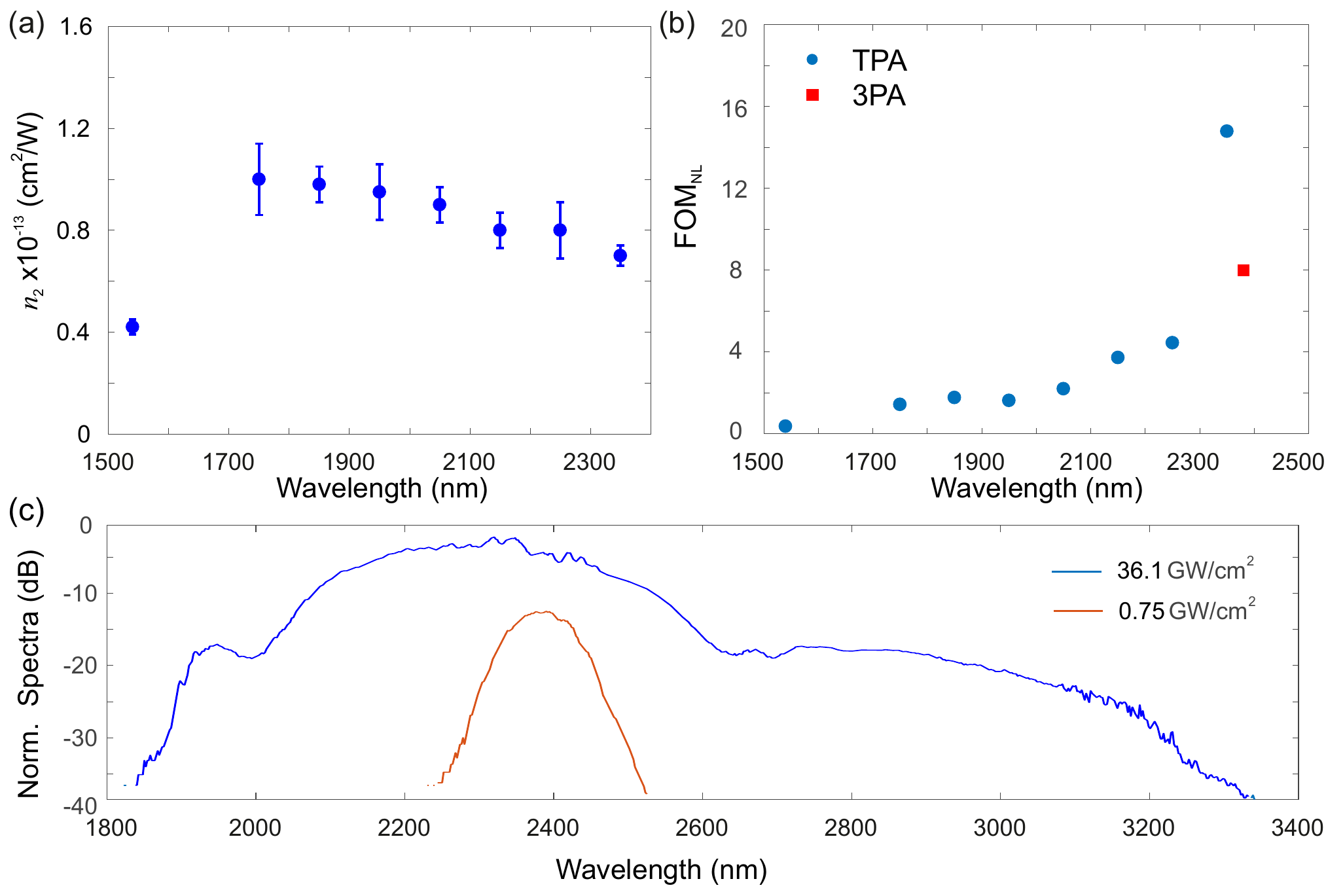}
\caption{(a) Wavelength dependence of the Kerr nonlinear coefficient $n_2$ for the tapered SCF. Error bars represent the uncertainty in the input pulses. (b) Wavelength dispersion of the FOM$_{{\rm NL}}$. (c) Continuum generated by pumping in the normal dispersion regime.}
\label{midfiber_fig4}
\end{figure}

The magnitude of the Kerr coefficient $n_2$ for each central wavelength is subsequently estimated by fitting the spectral broadening with the solutions to the full NLSE equation:

\begin{eqnarray}
 \frac{\partial A(z,t)}{\partial z} =  -\frac{i\beta_2}{2}\frac{\partial^2 A(z,t)}{\partial t^2}+i\gamma \vert A(z,t)\vert^2 A(z,t)-\frac{1}{2}(\alpha_l+\sigma_f)A(z,t) \label{NLSE3}
\end{eqnarray}
coupled to Eq. \ref{carrierdensity}. Here $A(z,t)$, $\beta_2$, and $\gamma$ represent the slowly varying pulse envelope, GVD, and nonlinear parameter, respectively. A complex nonlinear parameter is included to account for both the Kerr and TPA contributions: $\gamma=k_0n_2/A_{\rm{eff}}+i\beta_{\rm{TPA}}/2A_{\rm{eff}}$, where $A_{{\rm eff}}$ is the effective mode area. The corresponding values of $n_{2}$ are plotted in Fig. \ref{midfiber_fig4}(a), which show that as the input pulse wavelength shifts across the TPA edge, the $n_2$ value first increases slightly up to a value of $1.0\times 10^{-13} \,\rm{cm^2/W}$ at $1.75 \,\mu\rm{m}$, then drops to a modest value of  $0.7\times 10^{-13} \,\rm{cm^2/W}$ at $2.35 \,\mu\rm{m}$. This trend is consistent with the measurements by Bristow {\it et al.} \cite{bristow2007two} and is in good agreement with the nonlinear Kramers-Kr\"onig relation, where the values of $n_{2}$ peak at approximately the same wavelength. The measured Kerr coefficient $n_2$ of the polycrystalline SCF is of the same order of magnitude as previously reported for single crystal materials but slightly lower, which may be attributed to the different crystal orientations present in the cores \cite{zhang}. We note that the error bars represent experimental uncertainties originating from intensity fluctuations in the laser power and variations in the beam shape induced by tuning the OPO.

\subsection{Nonlinear figure of merit and continuum generation}

As a final step, the nonlinear parameters $\beta_{\rm{TPA}}$ and $n_{2}$ were used to investigate the dispersion of the FOM$_{{\rm NL}}$(defined as $n_{2}/\lambda\beta_{\rm{TPA}}$), plotted in Fig. \ref{midfiber_fig4}(b). This figure clearly shows that despite the decrease in $n_{2}$ at the longer wavelengths, the dramatic reduction in $\beta_{\rm{TPA}}$ results in a monotonic increase in the FOM$_{{\rm NL}}$. The value of the FOM$_{{\rm NL}}$ at $1.55\,\mu\rm{m}$ for this polycrystalline SCF is fairly modest $\sim 0.36$, though is comparable to previous reports in SOI waveguides \cite{lin2007two}, but the increasing trend indicates that these fibers should be suitable for nonlinear applications at wavelengths above $2\,\mu\rm{m}$. However, we note that the TPA definition for the FOM$_{{\rm NL}}$ is not entirely accurate for the final measurement at $2.35\,\mu\rm{m}$, as it is beyond the TPA edge, and a more accurate definition would be based on the 3PA parameter as, $n_{2}/\lambda\beta_{\rm{3PA}}I$. Using data for the 3PA parameter of single crystal silicon from the literature \cite {Gholami}, the 3PA FOM$_{{\rm NL}}$ is plotted as the red square in Fig. \ref{midfiber_fig4}(b) for comparison. Although this modified value is lower, it is clear that the FOM$_{{\rm NL}}$ continues its upward trend, so that one can expect higher nonlinear performance in the SCFs at longer wavelengths. %Our results show a complete picture of the FOM$_{{\rm NL}}$ in the silicon core material of the polycrystalline SCF as the wavelength moves into a regime of low nonlinear loss, thus indicate the potential of using the polycrystalline SCF platform for nonlinear applications when pumping near the TPA edge.

As a final demonstration of the optical quality of our polycrystalline SCFs, Fig. \ref{midfiber_fig4}(c) shows a supercontinuum spectrum generated when pumped with an input pulse at $\lambda_{p}=2.4\,\mu\rm{m}$ and an intensity of $36.1\,\rm{GW/cm^2}$. The bandwidth of the spectrum is $\sim 1600\,\rm{nm}$ at the $-40\,$dB level, corresponding to $\sim 0.9$ of an octave. Although this is not the broadest spectrum that has been generated in a crystalline silicon waveguide, to the best of our knowledge, it is the broadest that has been generated in the normal dispersion regime. Significantly, continuum spectra generated in the normal dispersion regime are often favoured for applications in spectroscopy and metrology as their flatness and coherence is better \cite{Heidt10}. This is because SPM dominates the broadening mechanism, so that phase noise introduced by processes such as four-wave mixing or modulation instability is kept to a minimum. When fully integrated with conventional mid-infrared fibers and components to improve the robustness of the system \cite{haonan}, we expect these SCFs could find wide-ranging applications in areas that require the generation and transmission of wavelengths in this important spectral band.

%by pumping this SCF further longer at the tuning edge of the OPO ($\lambda_{p}=2.4\,\mu\rm{m}$), a continuum up to $\sim 1600\,\rm{nm}$ ($\sim 0.9$ of an octave at -40 dB) can be generated with a moderate pump intensity ($36.1\,\rm{GW/cm^2}$), as shown in Fig. \ref{midfiber_fig4}(c). Although this pump wavelength is on the short wavelength side of the zero dipersion wavelength (ZDW), we attribute the large spectrum broadening is originating from small normal dispersion and very high FOM$_{{\rm NL}}$. In the literature, continua generated in the normal dispersion regime are not as broad as those generated in the anomalous dispersion regime, but the flatness and coherence of the continuum is better \cite{Heidt10}. This is because  SPM dominates the broadening mechanism and minimum the amplification of the phase noise in nonlinear process e.g, four-wave mixing or modulation instability. Nevertheless, the spectra, with the total spectral coverage extending from $1.8\,\mu\rm{m}$ to $3.4\,\mu\rm{m}$, which we believe to be the largest continuum generated in a silicon waveguide pumped in the normal dispersion region,  is likely to be of interest for applications in spectroscopy and metrology.

\section{Conclusion}

We have characterized both the linear and nonlinear transmission properties of our MCD SCFs from Telecom wavelengths up to the TPA edge in the mid-infrared regime. The dispersion curves obtained for the TPA and nonlinear refractive index are in good qualitative agreement with previous reports for single crystal silicon, indicating the high quality of the polysilicon core materials. The large spectral broadening measured for wavelengths beyond $2\,\mu$m suggest that polycrystalline SCFs are a viable platform for nonlinear applications within the short-wave and mid-infrared regimes, where applications include free-space communications, gas detection and medical diagnostics. We expect that continued efforts to understand the properties of these fibers and their highly nonlinear core material will help to establish their use in wide ranging areas of research.

\section*{Funding and Acknowledgments}

The authors wish to acknowledge financial support from the Engineering and Physical Sciences Research Council (EPSRC) (EP/P000940/1), National Natural Science Foundation of China (NNSFC) (Grant No.61705072), Natural Science Foundation of Hubei Province (2017CFB133), the Norwegian Research council, NORFAB and the J.E. Sirrine Foundation. All data supporting this study are openly available from the University of Southampton repository.

% \section*{Funding}
% Please identify all appropriate funding sources by name and contract number. Funding information should be listed in a separate block preceding any acknowledgments. List only the funding agencies and any associated grants or project numbers, as shown in the example below:\\
% \\
% National Science Foundation (NSF) (1253236, 0868895, 1222301); Program 973 (2014AA014402); Natural National Science Foundation (NSFC) (123456).\\
% \\
% OSA participates in \href{http://www.crossref.org/fundingdata/}{Crossref's Funding Data}, a service that provides a standard way to report funding sources for published scholarly research. To ensure consistency, please enter any funding agencies and contract numbers from the Funding section in Prism during submission or revisions.

% \section*{Acknowledgments}
% Acknowledgments, if included, should appear at the end of the document. The section title should not follow the numbering scheme of the paper.

% \section*{Disclosures}
% For \textit{Biomedical Optics Express} submissions only, disclosures should be listed in a separate nonnumbered section at the end of the manuscript. List the Disclosures codes identified on OSA's \href{http://www.osapublishing.org/submit/review/conflicts-interest-policy.cfm}{Conflict of Interest policy page}, as shown in the examples below:\\
% \\
% ABC: 123 Corporation (I,E,P), DEF: 456 Corporation (R,S). GHI: 789 Corporation (C).\\
% \\
% If there are no disclosures, then list ``The authors declare that there are no conflicts of interest related to this article.''


\begin{thebibliography}{99}

\bibitem{Leuthold2010} J. Leuthold, C. Koos, and W. Freude, ``Nonlinear silicon photonics,'' Nat. Photonics {\bf 4}(8), 535--544 (2010).

\bibitem{Ballato}   J. Ballato, T. Hawkins, P. Foy, R. Stolen, B. Kokuoz, M. Ellison, C. McMillen, J. Reppert, A. M. Rao, M. Daw, S. Sharma, R. Shori, O. Stafsudd, R. R. Rice, and D. R. Powers, ``Silicon optical fiber,'' \opex {\bf 16}(23), 18675-18683 (2008).


\bibitem{Orcutt}   J. S. Orcutt, S. D. Tang, S. Kramer, K. Mehta, H. Li, V. Stojanovi\'c, and R. J. Ram, ``Low-loss polysilicon waveguides fabricated in an emulated high-volume electronics process,'' \opex {\bf 20}(7), 7243-7254 (2012).

\bibitem{Peacock2016} A. C. Peacock, U. J. Gibson, and J. Ballato, ``Silicon optical fibres - past, present, and future,'' Adv. Phys.: X {\bf 1}, 114-127 (2016).

\bibitem{Nordstrand}  E. F. Nordstrand, A. N. Dibbs, A. J. Er{\aa}ker, and U. J. Gibson, ``Alkaline oxide interface modifiers for silicon fiber production,'' \ome{\bf 3}(5), 651-657 (2013).

\bibitem{Franz}  Y. Franz, A. F. J. Runge, H. Ren, N. Healy, K. Ignatyev, M. Jones, T. Hawkins, J. Ballato, U. J. Gibson, and A.
C. Peacock, ``Material properties of tapered crystalline silicon core fibers,'' \ome{\bf 7}(6), 2055-2061 (2017).


\bibitem{Fariza}  F. H. Suhailin, L. Shen, N. Healy, L. Xiao, M. Jones, T. Hawkins, J. Ballato, U. J. Gibson, and A. C. Peacock, ``Tapered polysilicon core fibers for nonlinear photonics,'' \ol {\bf 41}(7) 1360-1363 (2016).

%\bibitem{yin2007impact} L. Yin and G. P. Agrawal, ``Impact of two-photon absorption on self-phase modulation in silicon waveguides,'' \ol {\bf 32}, 2031--2033 (2007).

\bibitem{Liu2010} X. P. Liu, R. M. Osgood, Y. A. Vlasov, and W. M. J. Green,``Mid-IR optical parametric amplifier using
silicon nanophotonic waveguides,'' Nat. Photonics {\bf 4}(8), 557--560 (2010).

\bibitem{Zlatanovic2010} S. Zlatanovic, J. S. Park, S. Moro, J. M. C. Boggio, I. B. Divliansky, N. Alic, S. Mookherjea, and S. Radic, ``Mid-IR wavelength conversion in silicon waveguides using ultracompact telecom-band-derived pump source,'' Nat. Photonics {\bf 4}(8), 561--564 (2010).

\bibitem{bristow2007two} A. D. Bristow, N. Rotenberg, and  H. M. Van Driel, ``Two-photon absorption and Kerr coefficients of silicon for $850-2200\,\rm{nm}$,'' \apl {\bf 90}(19), 191104 (2007).

\bibitem{lin2007two} Q. Lin, J. Zhang, G. Piredda, R. W. Boyd, P. M. Fauchet, and G. P. Agrawal, ``Dispersion of silicon nonlinearities in the near infrared region,'' \apl {\bf 90}(2), 191104 (2007).

%\bibitem{Garcia} H. Garcia and R. Kalyanaraman,'' J. Phys. B {\bf 39}, 2737 (2010).

\bibitem{Liu2011nonlinear} X. P. Liu, J. B. Driscoll. J. I. Dadap, R. M. Osgood, S. Assefa, Y. A. Vlasov, and W. M. J. Green, ``Self-phase modulation and nonlinear loss in silicon nanophotonic wires near the mid-IR two-photon absorption edge,'' \opex {\bf 19}(8), 7778-7789 (2011).










% \bibitem{Duval}  S. Duval, M. Bernier, V. Fortin, J. Genest, M. Pich\'e, and R. Vall\'ee, ``Femtosecond fiber lasers reach the mid-IR,'' Optica {\bf 2}(7), 623-626 (2015).


\bibitem{peacock2012nonlinear} A. C. Peacock, P. Mehta, P. Horak, and N. Healy, ``Nonlinear pulse dynamics in multimode silicon core optical fibers,'' \ol {\bf 37}(16), 3351--3353 (2012).

\bibitem{Li80} H. H. Li, ``Refractive index of silicon and germanium and its wavelength and temperature derivatives,'' J. Phys. Chem. Ref. Data {\bf 9}, 561--658 (1980).

\bibitem{lagonigro2010low} L. Lagonigro, N. Healy, J. R. Sparks, N. F. Baril, P. J. A. Sazio, J. V. Badding, and A. C. Peacock, ``Low loss silicon fibers for photonics applications," \apl {\bf 96}(4), 041105 (2010).

\bibitem{Rouifed} M.-S. Rouifed, C. G. Littlejohns, G. X. Tina, Q. Haodong, T. Hu, Z. Zhang, C. Liu, G. T. Reed, and H. Wang, ``Low loss SOI waveguides and MMIs at the MIR wavelength of $2\,\mu\rm{m}$," IEEE Photonics Technol. Lett. {\bf 28}(24), 2827-2829 (2016).


\bibitem{Gholami} F. Gholami, S. Zlatanovic, A. Simic, L. Liu, D. Borlaug, N. Alic, M. Nezhad, Y. Fainman, and S. Radic, ``Third-order nonlinearity in silicon beyond $2350\,\rm{nm}$,'' \apl {\bf 99}(8), 081102 (2011).


\bibitem{mehta2010nonlinear} P. Mehta, N. Healy, N. F. Baril, P. J. A. Sazio, J. V. Badding, and A. C. Peacock, ``Nonlinear transmission properties of hydrogenated amorphous silicon core optical fibers,'' \opex {\bf 18}(16), 16826--16831 (2010).

\bibitem{Garcia2006} H. Garcia and R. Kalyanaraman,``Phonon-assisted two-photon absorption in the presence of a dc-field: the nonlinear Franz-Keldysh effect in indirect gap semiconductors,'' J. Phys. B {\bf 39}(12), 2737-2746 (2006).

\bibitem{Shen} L. Shen, N. Healy, P. Mehta, T. D. Day, J. R. Sparks, J. V. Badding, and A. C. Peacock, ``Nonlinear transmission properties of hydrogenated amorphous silicon core fibers towards the mid-infrared regime,'' \opex {\bf 21}(11), 13075--13083 (2013).




\bibitem{zhang} J. Zhang,``Anisotropic nonlinear response of silicon in the near-infrared region,'' \apl {\bf 91}(7), 071113 (2007).




\bibitem{Heidt10} A. M. Heidt, ``Pulse preserving flat-top supercontinuum generation in all-normal dispersion photonic crystal fibers,'' J. Opt. Soc. Am. B  {\bf 27}(3), 550-559 (2010).

 \bibitem{haonan}  H. Ren, O. Aktas, Y. Franz, A. F. J. Runge, T. Hawkins, J. Ballato, U. J. Gibson, and A. C. Peacock, ``Tapered silicon core fibers with nano-spikes for optical coupling via spliced silica fibers,'' \opex {\bf 25}(20), 24157-24163 (2017).

% \bibitem{Narayanan:10} K. Narayanan and S. F. Preble, ``Optical nonlinearities in hydrogenated-amorphous silicon waveguides,'' \opex {\bf 18}, 8998--9005 (2010).

%  \bibitem{grillet2012amorphous} C. Grillet, L. Carletti, C. Monat, P. Grosse, B. Ben Bakir, S. Menezo, J. M. Fedeli, and D. J. Moss, ``Amorphous silicon nanowires combining high nonlinearity, FOM and optical stability,'' \opex {\bf 20}, 22609--22615 (2012).

%  \bibitem{kuyken2011nonlinear} B. Kuyken, H. Ji, S. Clemmen, S. K. Selvaraja, H. Hu, M. Pu, M. Galili, P. Jeppesen, G. Morthier, S. Massar, L. K. Oxenl{\o}we, G. Roelkens, and R. Baets, ``Nonlinear properties of and nonlinear processing in hydrogenated amorphous silicon waveguides,'' \opex {\bf 19}, B146--B153 (2011).

%  \bibitem{mehta2010nonlinear} P. Mehta, N. Healy, N. F. Baril, P. J. A. Sazio, J. V. Badding, and A. C. Peacock, ``Nonlinear transmission properties of hydrogenated amorphous silicon core optical fibers,'' \opex {\bf 18}, 16826--16831 (2010).

%  \bibitem{Narayanan:1018} K. Narayanan, A. W. Elshaari, and S. F. Preble, ``Broadband all-optical modulation in hydrogenated-amorphous silicon waveguides,'' \opex {\bf 18}, 9809--9814 (2010).

% \bibitem{mehta2011all} P. Mehta, N. Healy, T. D. Day, J. R. Sparks, P. J. A. Sazio, J. V. Badding, and  A. C. Peacock, ``All-optical modulation using two-photon absorption in silicon core optical fibers,'' \opex {\bf 19}, 19078--19083 (2011).

% \bibitem{Clemmen10} S. Clemmen, A. Perret, S. K. Selvaraja, W. Bogaerts, D. van Thourhout, R. Baets, Ph. Emplit, and S. Massar, ``Generation of correlated photons in hydrogenated amorphous-silicon waveguides," \ol {\bf 35}, 3483--3485 (2010).

% \bibitem{kuyken2011chip} B. Kuyken, S. Clemmen, S. K. Selvaraja, W. Bogaerts, D. Van Thourhout, Ph. Emplit, S. Massar, G. Roelkens, and R. Baets, ``On-chip parametric amplification with 26.5 dB gain at telecommunication wavelengths using CMOS-compatible hydrogenated amorphous silicon waveguides," \ol {\bf 36}, 552--554 (2011).

% \bibitem{Wang12} K.-Y. Wang and A. C. Foster, ``Ultralow power continuous-wave frequency conversion in hydrogenated amorphous silicon waveguides," \ol {\bf 37}, 1331--1333 (2012).

% \bibitem{Kuyken11} B. Kuyken, X. Liu, G. Roelkens, R. Baets, R. M. Osgood, Jr., and W. M. J. Green, ``$50\,$dB parametric on-chip gain in silicon photonic wires," \ol {\bf 36}, 4401--4403 (2011).

% \bibitem{Kuyken11b} B. Kuyken, X. Liu, R. M. Osgood Jr., R. Baets, G. Roelkens, and W. M. J. Green, ``mid-IR to telecom-band supercontinuum generation in highly nonlinear silicon-on-insulator wire waveguides,'' \opex {\bf 19}, 20172--20181 (2011).

% \bibitem{baril2011confined} N. F. Baril, R. He, Rongrui, T. D. Day, J. R. Sparks, B. Keshavarzi, M. Krishnamurthi, A. Borhan, V.  Gopalan, A. C. Peacock, N. Healy, P. J. A. Sazio, and J. V. Badding, ``Confined high-pressure chemical deposition of hydrogenated amorphous silicon,'' J. Am. Chem. Soc. {\bf 134}, 19--22 (2011).

% \bibitem{Ballato10} J. Ballato, T. Hawkins, P. Foy, B. Yazgan-Kokuoz, C. McMillen, L. Burka, S. Morris, R. Stolen, and R. Rice, ``Advancements in semiconductor core optical fiber,'' Opt. Fiber Technol. {\bf 16}, 399--408 (2010).

% \bibitem{mehta2012ultrafast} P. Mehta, N. Healy, T. D. Day, J. V. Badding, and A. C. Peacock, ``Ultrafast wavelength conversion via cross-phase modulation in hydrogenated amorphous silicon optical fibers,'' \opex {\bf 20}, 26110--26116 (2012).

% \bibitem{Healy10} N. Healy, J. R. Sparks, P. J. A. Sazio, J. V. Badding, and A. C. Peacock, ``Tapered silicon optical fibers,'' \opex {\bf 18}, 7596--7601 (2010).

% \bibitem{yin2007impact} L. Yin and G. P. Agrawal, ``Impact of two-photon absorption on self-phase modulation in silicon waveguides," \ol {\bf 32}, 2031--2033 (2007).

% \bibitem{Li80} H. H. Li, ``Refractive index of silicon and germanium and its wavelength and temperature derivatives," J. Phys. Chem. Ref. Data {\bf 9}, 561--658 (1980).

% \bibitem{Matres13} J. Matres, G. C. Ballesteros, P. Gautier, J.-M. F\'{e}d\'{e}li, J. Mart\'{\i}, and C. J. Oton, ``High nonlinear figure-of-merit amorphous silicon waveguides,'' \opex {\bf 21}, 3932--3940 (2013).

% \bibitem{peacock2012nonlinear} A. C. Peacock, P. Mehta, P. Horak, and N. Healy, ``Nonlinear pulse dynamics in multimode silicon core optical fibers," \ol {\bf 37}, 3351--3353 (2012).



% \bibitem{selvaraja2009low} S. K. Selvaraja, E. Sleeckx, M. Schaekers, W. Bogaerts, D. Van Thourhout, P. Dumon, and R. Baets, ``Low-loss amorphous silicon-on-insulator technology for photonic integrated circuitry," Opt. Commun. {\bf 282}, 1767--1770 (2009).

% \bibitem{shoji2010ultrafast} Y. Shoji, T. Ogasawara, T. Kamei, Y. Sakakibara, S. Suda, K. Kintaka, H. Kawashima, M. Okano, T. Hasama, H. Ishikawa, and M. Mori, ``Ultrafast nonlinear effects in hydrogenated amorphous silicon wire waveguide," \opex {\bf 18}, 5668--5673 (2010).

% \bibitem{Sanghera06} J. S. Sanghera, I. D. Aggarwal, L. B. Shaw, C. M. Florea, P. Pureza, V. Q. Nguyen, F. Kung, and I. D. Aggarwal, ``Nonlinear properties of chalcogenide glass fibers," J. Optoelectron. Adv. M. {\bf 8}, 2148--2155 (2006).



\end{thebibliography}
\end{document}